\documentstyle[preprint,pre,aps,eqsecnum]{revtex}

\def\beq{\begin{equation}}
\def\eeq{\end{equation}}
\def\beqn{\begin{eqnarray}}
\def\eeqn{\end{eqnarray}}
\def\ra{\rightarrow}

\begin{document}
\preprint{IFUP-TH 31/94}
\draft
\title{Some properties of renormalons in gauge theories}
\author{G. Di Cecio and  G. Paffuti}
\address{Dipartimento di Fisica dell'Universit\`a  and
I.N.F.N., sez. di Pisa\\
I-56100 Pisa, Italy}
\maketitle
\begin{abstract}
We find the explicit operatorial form of renormalon-type singularities in
abelian gauge theory. Local operators of dimension six take care of the first 
U.V. renormalon, non local operators are needed for I.R. singularities.
In the effective lagrangian constructed with these operators non local
imaginary parts appearing in the usual perturbative expansion  at large    
orders are cancelled.
\end{abstract}
\thispagestyle{empty}
\newpage
\section{Introduction}
It is well known  that in  field theory the perturbative series
diverge \cite{Dyson} and are  asymptotic at best so that the question of the
reliability of the perturbative expansion is raised. In particular the
factorial growth of the perturbative coefficients has been related to
the existence of instanton solutions \cite{Lipatov} and, in this way, 
 a great number of results has been found (for a review see
\cite{Largeorder}). 
The perturbative expansion for
renormalizable  field theories is plagued
by further divergences, the so-called renormalons, which show up as
singularities on the real axis of the Borel transform of the
perturbative series \cite{tHooft,Lautrup}. Moreover this kind of
problems has recently gained a great interest
\cite{West,Brown1,Brown2,Zakharov,BenZak,Benekepl,Grunberg,Muellerpl} and many
questions, mainly concerning the asymptotic behaviour of 
perturbative expansions in QCD, still wait for an answer. 

The standard technique to cope with divergent series is the
introduction of the Borel transform. If a function $F(\alpha)$ admits
an asymptotic expansion and satisfies certain analytic properties
\cite{Watson,Sokal}, it can be uniquely reconstructed from the Borel
transform $\tilde{F}(b)$ defined by:
\beq
F(\alpha )=\int^\infty_0db\,{\rm e}^{-b/\alpha }\tilde{F}(b)\label{borel}
\eeq
The presence of singularities (poles or cuts) on the positive real
axis of the Borel transform forces  to give a contour prescription
(for example a principal value prescription) in order to define the
perturbative expansion via the integral (\ref{borel}). This results in
an ambiguity in the sum of the perturbative series given by the
integration above and below the positive real axis in the Borel
plane.

\noindent 
To be definite, a singularity of the form:
\beq
\tilde{F}(b)\sim\frac{1}{(b-b_0)^{1-s}}
\eeq
gives rise to an imaginary part in the function $F(\alpha )$ of the form:
\beq
{\rm Im}F(\alpha )\sim \alpha^s{\rm e}^{-b_0/\alpha }\label{im}
\eeq
which in turn is related, via a dispersion relation, to the large
order behaviour of the coefficients $f_n$ of the perturbative series:
\beq
f_n\mathop \sim_{n\ra\infty } const\left(\frac{1}{b_0}\right
)^n\,n^s\,\Gamma (n-s)\label{largeorder}
\eeq
The renormalon-type singularities occurr for 
\beq
b_0=\pm n/\beta_2\label{poles}
\eeq
where $n$ is a positive integer and $\beta_2$ is the first term of the
$\beta$-function and the sign plus refers to the ultraviolet renormalons
while the sign minus to the  infrared ones. These divergences are
usually the closest ones to the
origin and  give the leading asymptotics for the perturbative series.

In the following we want to study the structure and a possible
classification of the singularities due to renormalons, in the
framework of QED with a large number of flavours \cite{Coq}.

The aim of the first part of the paper is to present a prescription
which could allow to take sistematically into account the imaginary
parts (see eq. (\ref{im})) coming from ultraviolet renormalons. The
idea stems from a conjecture made for the first time by Parisi for the
$\phi^4$ theory \cite{Parisi1} and explicitly checked in the $1/N$
limit of this theory \cite{Parisi1,Bergere,preprint}.

This conjecture relates the imaginary parts coming from the renormalon
at $b=n/\beta_2$ to the insertion at zero momentum of local operators
of dimension $d=2n+4$. Renormalization group arguments fix the
form of the imaginary parts from which the asymptotic behaviour
(\ref{largeorder}) is extracted (with given $b_0$ and $s$) up to an
overall normalization constant.

We implement this conjecture adding gauge-invariant operators as
imaginary counterterms in the lagrangian. The feasibility of such a
description for the renormalon-type singularities relies on the
fulfilment of two conditions:
\begin{itemize}
\item A unique choice must exist for the operators in order to remove the
imaginary parts of all Green functions.
\item In the effective theory described by the lagrangian with the
counterterms, non local singularities in the Borel transformed Green
functions must be absent.  
\end{itemize}

The second half of the paper concerns the problem of infrared
renormalons \cite{Parisi2,David,Muellernp}. In the case of QCD a
connection  has
been conjectured  between infared renormalons and the
admissible multilocal gauge-invariant operators which can be seen as
the dual of the local operators of the theory \cite{Parisi2}. Then
from the absence of local gauge-invariant operators of dimension two,
the absence of the infrared renormalon at
$b=-1/\beta_2$ has been argued.
Recently the possibility of the existence of this
renormalon has been raised  \cite{Brown1,Brown2} and this would
suggest  the presence of
a $1/q^2$ term in the operator product expansion for euclidean
correlation functions \cite{Shifman} and would spoil the connection
with the gauge-invariant operators.

While on general grounds this possibility cannot be ruled out, an
explicit calculation \cite{Benekenp} shows that this pole 
doesn't indeed exist within the $1/N_f$ expansion for QED. Moreover QCD is
believed to have the same behaviour of QED after replacing the right
$\beta_2$ coefficient.

In the spirit of the calculation of \cite{Benekenp} we  study
the structure of the first infrared renormalons in QED. In particular
we find the form of the multilocal counterterms
compensating the imaginary parts arising from infrared renormalons
and, in connection with the case of the ultraviolet renormalons, we 
check that a unique prescription can be given.

A short comment is in order about the use of the $1/N_f$ expansion.
This kind of expansion is the natural framework for the study of
renormalons in that it provides a graphic tool (see figures) for the
investigation of these problems\footnote{After completion of this
work, we came at vision of \cite{Vainshtein} in which the authors
conclude that it is not possible in general to associate the
ultraviolet renormalons to a well-defined set of graphs; this is
indeed possible in the large $N_f$ limit; on the other hand the 
calculations performed in our paper are at fixed order in
$1/N_f$.}. Moreover it allows a partial resummation of the
perturbative series so that one can obtain informations on the whole
series  and calculate quantities, like the overall normalization
constant in (\ref{largeorder}), which cannot be found at any finite
order in perturbation theory \cite{Grunberg}.

The paper is organized as follows. In Sect.II we introduce the
notations and study the problem of the first ultraviolet renormalon.  
We find the explicit form of the counterterms and we check that all
singularities can be absorbed in such a way. Moreover we verify that
the non local singularities which appear in the naive perturbation
theory are indeed absent when considering the lagrangian with the
counterterms. In Sect.III we perform similar
calculations for the infrared renormalons; the main difference is
that, in this case, the operators to be added to the lagrangian are
multilocal. Sec.IV is devoted to some considerations about the meaning
of the lagrangian with the imaginary counterterms. Finally, the
Appendix contains some technical details about the calculations with the
background field formalism.

\section{Ultraviolet renormalons}

Let us begin this section with the introduction of some definitions
about the $1/N_f$ expansion in QED.

Defining a rescaled coupling constant $a=\alpha N_f$, the $1/N_f$
expansion is generated, at fixed $a$, by considering a factor of
$N_f^{-1/2}$ for every interaction vertex and a factor of $N_f$ for
every closed fermion loop. Then the photon propagator is given at
leading order by the sum of the bubble diagrams (fig.1). In the Landau
gauge it reads: 
\beq
D_{\mu\nu}(k)=-\frac{k_\mu k_\nu -k^2 g_{\mu\nu}}{k^4}D(k)
\eeq
where:
\beq
D(k)=\frac{1}{1+\Pi_0 (k)}\label{phprop}
\eeq
and $\Pi_0 (k)$ is the single bubble diagram. In the following we are
interested in the high momentum region of integration in the Feynman
integrals; so we need the expression of $\Pi_0 (k)$ in the deep
euclidean region. After renormalization in the $\overline{{\rm MS}}$ scheme,
we have (all fermions are assumed to have the same mass):  
\beq
\Pi_0 (k) \mathop \simeq_{k\gg m}-\frac{a}{3\pi}\left \{ {\rm log }
\frac{k^2}{\mu^2} -\frac{5}{3}\right \} + \frac{2a}{\pi}\frac{m^2}{k^2} +
 O\left (\frac{1}{k^4}\right )
\eeq
From (\ref{phprop}) and the definition (\ref{borel}) one can easily
derive the expression for the Borel transform, with respect to $a$, of
the photon propagator; introducing an extra factor of $a$, we have: 
\beq
F(k,b)\equiv \widetilde{(aD)}(k,b)\mathop \simeq_{k\gg m} {\rm e}^{-5b/9\pi}\left
(\frac{k^2}{\mu^2}\right )^{b/3\pi}\left [1-\frac{2bm^2}{\pi k^2}+O\left (\frac{1}{k^4}\right )\right ]\label{effe}
\eeq

\subsection{Imaginary part of Green functions}
At next-to-leading order in $1/N_f$, the Green functions correspond to
the graphs with one insertion of the chain of the bubbles (fig.2-4)
and present renormalon-type singularities. In fact the Feynman
integrals, in which the photon propagator (\ref{phprop}) appears in the
integrand, have an ambiguity arising from the region of integration
around the Landau pole (i.e. the pole in (\ref{phprop})) which occurr,
for infrared free theories (in our case $\beta (a)=a^2/3\pi +O(1/N_f)$)
in the ultraviolet region:
\beq
\Lambda^2=\mu^2{\rm e}^{5/3} {\rm e}^{1/\beta_2 a}\label{lanpole}
\eeq
(we  are always assuming the $\overline{{\rm MS}}$ scheme). This
ambiguity causes the Green functions to develop an imaginary part and
is related to the leading factorial growth of the coefficients of the
perturbative series in $\alpha$. In fact it turns out that, after
expanding (\ref{phprop}) in powers  of $\alpha$, the coefficient of
order $n$ receives a contribution proportional to $n!$ from the
ultraviolet region of integration. These problems appear as poles on
the real positive axis of the Borel transformed Green functions.

In the following we intend to find the behaviour of the Borel
transform of the vertex function, of the electron propagator and of
the photon vacuum polarization near the first ultraviolet renormalon at
$b=3\pi$. This will enable us to find the imaginary part of these
functions and to study the connection with local gauge-invariant
operators of dimension six.

Let us begin with the Borel transform of the vertex function (fig.2).
In the euclidean space the term of order $1/N_f$ is given by:
\beq
\widetilde{\left (\frac{\Gamma_\mu}{e}\right )}
 = -\frac{4\pi}{N_f} \int \frac{d^4k}{(2\pi )^4}\,\frac{\gamma_\rho
[\hat{p^\prime}-\hat{k}-m]\gamma_\mu
[\hat{p}-\hat{k}-m]\gamma_\sigma}{[(p^\prime
-k)^2+m^2][(p-k)^2+m^2]}\,\frac{k_\rho k_\sigma -k^2\delta_{\rho\sigma}}{k^4}\,F(k,b)\label{gamma}
\eeq
where we have factorized out the electron charge $e$. Renormalization
is taken into account by simply introducing the renormalized bubble.
In fact, since we are working in the Landau gauge, the vertex
function needs no further renormalization and this results in the
absence of a pole in $b=0$ in $\tilde{\Gamma}_\mu$ as written in
(\ref{gamma}). Moreover introducing the term of order $1/k^{2n}$ in
the expansion (\ref{effe}) of $F(k,b)$ results in a pole in
$\tilde{\Gamma}_\mu$ at $b=3\pi (n+1)$.

After some manipulations with $\gamma$ matrices and performing the
integral we  obtain for the first pole:
\beqn
\widetilde{\left (\frac{\Gamma_\mu}{e}\right )}\;\;\mathop 
\simeq _{b\sim 3\pi}\frac{1}{N_f}\frac{1}{8}\left (\frac{1}{\mu^2}\right )^{b/3\pi}\frac{{\rm e}^{-5b/9\pi}}{b-3\pi}\{[(p-p^\prime )^2-9m^2-6p\cdot p^\prime ]\gamma_\mu\nonumber\\ 
+3\hat{p^\prime}\gamma_\mu \hat{p}-4(p^\prime_\mu
-p_\mu)(\hat{p^\prime}-\hat{p})\}\;\;\;\;\;\;\;\;\;\;\;\;\;\;\;\;\;\;\;\;\;\;\;\;\;\;\;\;\;\;\;\;\;\label{ver}
\eeqn 
From the Borel representation (\ref{borel}) we deduce that
$\Gamma_\mu$ develops an imaginary part given by the residue at the
pole in $b=3\pi$:
\beq
{\rm Im}\;\Gamma_\mu =\pi e\; {\rm Res}\left\{\left
(\widetilde{\frac{\Gamma_\mu}{e}}\right ) {\rm e}^{-b/a}\right \}_{b=3\pi}
\eeq
Then (\ref{ver}) yields:
\beq
{\rm Im}\;\Gamma_\mu =\frac{1}{N_f}\frac{\pi e}{8\Lambda^2}\left\{[4q^2-3p^2-3p^{\prime 2}-9m^2]\gamma_\mu + 3\hat{p^\prime}\gamma_\mu\hat{p}-4q_\mu\hat{q}\right\} + O\left(\frac{1}{\Lambda^4}\right )\label{imver}
\eeq
On the mass-shell this expression reduces to:
\beq
{\rm Im}\;\Gamma_\mu =\frac{1}{N_f}\frac{\pi e}{2\Lambda^2}q^2\gamma_\mu
\eeq
hence there is no contribution of the first ultraviolet renormalon to
the large order behaviour of the perturbative series of the anomalous
magnetic moment of the electron, in agreement with the asymptotic
behaviour found in \cite{Lautrup}.
  
From (\ref{imver}) we see that  the imaginary parts arising from
different  renormalons are
classified, at this order in $1/N_f$, in terms of the renormalization
group invariant (i.e. $\mu$-independent) $\Lambda$-parameter which
plays the role of a dimensional expansion parameter. It must be noted
that this parameter encloses all the dependence of the calculation
from the choice of the renormalization scheme. 

Similar considerations hold for the fermion propagator (fig.3). Its
Borel transform at order $1/N_f$ reads:
\beq
\tilde{\Sigma}(p,b)=\frac{4\pi}{N_f}\int\frac{d^4k}{(2\pi
)^4}\,\frac{\gamma_\mu
[(\hat{p}+\hat{k})-m]\gamma_\nu}{(p+k)^2+m^2}\,\frac{k^2\delta_{\mu\nu}-k_\mu
k_\nu }{k^4}\, F(k,b)
\eeq
Isolating the pole at $b=3\pi$ we have:
\beq
\tilde{\Sigma}(p,b)\mathop \simeq_{b\sim 3\pi}\frac{1}{N_f}\left
(\frac{1}{\mu^2}\right )^{b/3\pi}\frac{{\rm
e}^{-5b/9\pi}}{b-3\pi}\left \{ -\frac{9}{8}m^2\hat{p}+\frac{9}{4}m^3-\frac{3}{8}p^2\hat{p}+\frac{9}{2}\frac{bm^3}{\pi}\right\}
\eeq
This yields an imaginary part given by:
\beq
{\rm Im}\;\Sigma (p) =\frac{1}{N_f} \frac{\pi}{\Lambda^2}\left\{-\frac{9}{8}m^2\hat{p}-\frac{3}{8}p^2\hat{p}+\frac{63}{4}m^3\right\}+O\left(\frac{1}{\Lambda^4}\right)\label{imself}
\eeq 
It is straightforward to verify that (\ref{imver}) and (\ref{imself})
satisfy the Ward identity:
\beq
q_\mu\, {\rm Im}\,\left (\frac{\Gamma_\mu}{e}\right )= {\rm Im}\,\Sigma
(p^\prime )-{\rm Im}\,\Sigma (p)\label{ward}
\eeq

Finally, the imaginary part of the photon vacuum polarization can be
deduced from Beneke \cite{Benekenp}. Exploiting gauge-invariance for
the proper photon prapagator we have:
\beq
\Pi_{\mu\nu}(q^2)=(q_\mu q_\nu -q^2g_{\mu\nu}) \Pi (q^2)
\eeq
with:
\beq
\Pi =\Pi_0 +\Pi_1\frac{1}{N_f}+O\left (\frac{1}{N_f^2}\right )
\eeq
$\Pi_0$ is the single bubble, while $\Pi_1$ corresponds to the
diagrams of fig.4. The result found by Beneke in the massless
limit is:
\beq
\widetilde{\left (\frac{\Pi_1}{a}\right )}(Q^2,b)=-\frac{1}{6\pi^2}{\rm e}^{-5b/9\pi}\left (\frac{Q^2}{\mu^2}\right )^{b/3\pi}D(b/3\pi )
\eeq
with:
\beq
D(u)\mathop \simeq_{u\ra 1} \frac{2}{3}\frac{1}{(1-u)^2}+\frac{11}{9}\frac{1}{1-u} + O(1)
\eeq   
From this expression one can infer the imaginary part of the proper photon propagator:
\beqn
{\rm Im}\;\Pi_{1\,\mu\nu}(q^2)=&-&\frac{a}{3}\frac{q^2}{\Lambda^2}(q_\mu q_\nu -q^2g_{\mu\nu})\left ({\rm log}\frac{q^2}{\mu^2}-\frac{5}{3}\right )\nonumber\\
&+&\frac{q^2}{\Lambda^2}(q_\mu q_\nu -q^2g_{\mu\nu})\left
(\frac{11}{18}a + \pi \right )+O\left (\frac{1}{\Lambda^4}\right ) \label{impol}
\eeqn 
The non local singularity corresponding to the term proportional to
${\rm log}\, q^2$ in the above formula could in principle destroy the
connection with the gauge-invariant counterterms. Its role will be
discussed in the following.

\subsection{Connection with local operators of dimension six}
Our basic result is the explicit value of the imaginary parts of the
vertex function (eq.(\ref{imver})), of the fermion propagator
(eq.(\ref{imself})) and of the photon vacuum polarization
(eq.(\ref{impol})) in correspondence of the first ultraviolet
renormalon. We want to stress again that all our calculations have
been performed at fixed order in $1/N_f$ and all the following
considerations rely upon this kind of expansion in which
renormalon-type problems have a more direct interpretation.

Let us now turn to the main point of our investigation i.e. the
connection with local gauge-invariant operators. On simply dimensional
grounds and exploiting renormalization group arguments one can argue
that the imaginary parts given by (\ref{imver}), (\ref{imself}) and
(\ref{impol}) are proportional to the insertion of operators of
dimension six. We want to check if it is possible to give a
prescription in terms of local gauge-invariant operators added as
imaginary counterterms in the lagrangian. Let us study the structure
of the counterterms.

The term proportional to $(q^2\gamma_\mu -q_\mu \hat{q})$ in
(\ref{imver}) can be seen as the insertion of the gauge-invariant operator   
$(\partial_\mu {\rm F}_{\mu\nu})\bar{\psi}\gamma_\nu\psi$; hence we
deduce the presence of the following counterterm in the lagrangian
(the coefficient is uniqely determined):
\beq
\Delta {\cal L}_1=-i\frac{1}{N_f}\frac{\pi e}{2\Lambda^2}(\partial_\mu {\rm F}_{\mu\nu})\bar{\psi}\gamma_\nu\psi \label{delta1}
\eeq
Moreover, from the term proportional to $m^2$ in (\ref{imver}) and
(\ref{imself})  we see that the operators $m^2\Lambda^{-2}
\bar{\psi}\hat{\partial}\psi$  and $em^2\Lambda^{-2}
\bar{\psi}\hat{A}\psi$ combine  
in a gauge-invariant form and this corresponds to the following
lagrangian  counterterm:
\beq
\Delta{\cal L}_2=-i\frac{1}{N_f}\frac{9\pi
}{8}\frac{m^2}{\Lambda^2}\bar{\psi}(i\hat{\partial} - e\hat{{\rm A}})\psi\label{delta2}
\eeq

The fundamental check involves the imaginary part of the photon vacuum
polarization which contains a term proportional to log$\,q^2$. In fact
this term cannot be cancelled by the direct insertion of a local
operator but ought to arise from the insertion of a local operator in
a loop diagram. One can easily convince himself that the only operator
that can do this job must have the form of (\ref{delta1}). We then
compute the insertion of $\Delta {\cal L}_1$ on the vacuum
polarization as given by the diagram of fig.5. This diagram is of
order $1/N_f$ since it contains a term $1/N_f$ from $\Delta {\cal
L}_1$, a term $1/N_f$ from the two vertices and a term $N_f$ from the
fermion loop. It gives the following contribution to the imaginary
part of the vacuum polarization (in the $\overline{{\rm MS}}$ scheme):
\beq
i\frac{1}{N_f}\frac{a}{3}\frac{q^2}{\Lambda^2}(q_\mu q_\nu -q^2g_{\mu\nu})\left ({\rm log}\frac{q^2}{\mu^2}-\frac{5}{3}\right )
\eeq
This contribution exactly cancels the logarithmic term in Im$\,\Pi_1$.

To summarize, we have explicitly checked that, at next-to-leading
order in $1/N_f$, the imaginary parts due to the ultraviolet
renormalon at $b=3\pi$ can be cancelled by   the contribution of local
gauge-invariant operators of dimension six. Indeed it is possible to
give a realization of this connection at lagrangian level.

\section{Infrared renormalons}

In this section we deal with the so-called infrared renormalons i.e.
singularities on the real axis of the Borel transform at $b\beta_2=-n$
which arise from the low momentum region of integration  in the Feynman
integrals. As already stated these singularities are on the negative
real axis of the Borel variable for infrared free theories like QED or
QCD in the limit of large number of flavours and give rise to a sign
alternating Borel summable contribution to the perturbative series
(see eq.(\ref{largeorder})). 

On the contrary, for asymptotically free theories like QCD the
 infrared renormalons destroy the Borel summability of the
perturbative series, a phenomenon usually related to the presence of
non-perturbative effects, the condensates, which cause the physical
vacuum to have a non trivial structure. Hence the study of infrared
renormalons in QCD is of primary interest expecially for the
comprehension of the operator product expansion which is usually
introduced in order to parametrize non-perturbative effects
\cite{Shifman}.

Neverthless it is common wisdom that the structure of the infrared
renormalons in QCD can be obtained from the study of QED, after
the substitution of the value of $\beta_2$ of QCD. This means that the
structure of the infrared renormalons is insensitive to the presence
of non-perturbative contributions in the operator product expansion
which in the case of QED is a pure reorganization of the
perturbative series but depends only on the nature of the
gauge-invariant operators of the theory, in agreement with the
conjecture made by Parisi \cite{Parisi2}.

In the following we intend to perform an explicit check of the
connection with gauge-invariant operators, in analogy with the
calculations of the previous section concerning ultraviolet
renormalons. The basic difference in the case of infrared renormalons
is that this connection can be implemented by adding imaginary
multilocal gauge-invariant operators to the lagrangian. These
multilocal operators must be considered as the ``dual'' of the local
operators appearing in the operator product expansion (for a rigorous
definition see \cite{Parisi2}). Our aim is to find the form and the
normalization of these operators and to check the combinatoric.

In analogy with the preceding section we deal with QED at
next-to-leading order in $1/N_f$. The only difference is that now we
consider the massless case; then the proper photon propagator at
leading order has the following form in the  $\overline{{\rm MS}}$
scheme: 
\beq
\Pi_0 (k) =-\frac{a}{3\pi}\left ({\rm log}\frac{k^2}{\mu^2}-\frac{5}{3}\right )
\eeq
while the  Borel transform of $D(k)$ is:
\beq
F(k,b)={\rm e}^{-5b/9\pi}\left (\frac{k^2}{\mu^2}\right
)^{b/3\pi}\label{massless} 
\eeq
The expression for the Borel transformed functions is the same as
before. From (\ref{gamma}) and (\ref{massless}) we see that
$\tilde{\Gamma}_\mu$ has infrared singularities for $b=-3\pi n$;
expanding around the first two poles we have:
\beqn
\widetilde{\left (\frac{\Gamma_\mu}{e}\right )}\simeq\frac{1}{N_f}\left
(\frac{1}{\mu^2}\right )^{b/3\pi}{\rm
e}^{-5b/9\pi}\left\{\frac{9}{8}\,\frac{1}{b+3\pi}\,
\frac{1}{\hat{p}}\gamma_\mu\frac{1}{\hat{p}^\prime}-\frac{1}{4}\,\frac{1}{b+6\pi}
\,\right [\frac{2q^2}{p^2p^{\prime 2}}\,\hat{p}\gamma_\mu\hat{p^\prime}+\nonumber\\
+\left. \left.  3\gamma_\mu-5
\frac{p^\prime_\mu\hat{p^\prime }}{p^{\prime
2}}-5\frac{p_\mu\hat{p}}{p^{ 2}}-\frac{p_\mu\hat{p^\prime }}{p^{\prime
2}}-\frac{p^\prime_\mu\hat{p }}{p^{ 2}}\right ]\right \} 
\eeqn
This yields an imaginary part given by:
\beqn
{\rm Im}\;\Gamma_\mu=\frac{1}{N_f}\frac{9\pi e}{8}\Lambda^2\frac{1}{\hat{p}}\gamma_\mu\frac{1}{\hat{p^\prime}}
-\frac{1}{N_f}\frac{\pi e}{4}\frac{\Lambda^4}{p^2p^{\prime 2}}\left
(\frac{2q^2}{p^2p^{\prime
2}}\hat{p}\gamma_\mu\hat{p^\prime}+3\gamma_\mu -\right.\nonumber\\ 
\left. -5\frac{p^\prime_\mu\hat{p^\prime }}{p^{\prime
2}}-5\frac{p_\mu\hat{p}}{p^{ 2}}-\frac{p_\mu\hat{p^\prime }}{p^{\prime 2}}-\frac{p^\prime_\mu\hat{p }}{p^{ 2}}\right )+O(\Lambda^6)\label{imverir}
\eeqn
While for the ultraviolet renormalons the expansion parameter was
$\Lambda^{-2}$, now it is $\Lambda^2$; in fact we are simulating
ultraviolet free theories which present a Landau pole at small momenta
(in (\ref{lanpole}) we must take $\beta_2 <0$)

The imaginary part of the fermion propagator can be obtained in a
similar way. We have:
\beq
{\rm Im}\;\Sigma (p)=\frac{1}{N_f}\frac{9\pi
}{8}\frac{\Lambda^2}{p^2}\hat{p}-\frac{1}{N_f}\frac{3\pi}{4}\frac{\Lambda^4}{p^4}\hat{p}+O(\Lambda^6)\label{imselfir}
\eeq
Even (\ref{imverir}) and (\ref{imselfir}) satisfy the Ward identity
(\ref{ward}). 

\noindent
From \cite{Benekenp} we deduce the following imaginary part for the
photon vacuum polarization:
\beq
{\rm Im}\;\Pi_{1\,\mu\nu} = \frac{3a}{4}\frac{\Lambda^4}{q^4}(q_\mu
q_\nu -q^2g_{\mu\nu})+O(\Lambda^6)\label{impolir}
\eeq
The absence of the term of order $\Lambda^2$ reflects the absence of
the first renormalon pole at $b=3\pi$ and this is a first evidence for
the connection with gauge-invariant operators
\cite{Brown1,Brown2,Benekenp}. 

We want to study how the imaginary parts (\ref{imverir}),
(\ref{imselfir}) and (\ref{impolir}) can be compensated by the
contributions of multilocal operators.

First of all it is evident that the terms of order $\Lambda^2$ in
(\ref{imverir}) and in (\ref{imselfir}) are both proportional to the
insertion on the functions calculated at leading order (without the
chain of bubbles) of the multilocal operator:
\beq
\Delta S_1=-i\frac{1}{N_f}\frac{9\pi}{16}\Lambda^2\int d^4x\bar{\psi}\gamma_\mu\psi(x)\int d^4y\bar{\psi}\gamma_\mu\psi(y)\label{deltas1}
\eeq
It is pleasant that a unique choice of the normalization of the
operator takes care of both the contributions.

Let us now turn to eq.(\ref{impolir}). When considering the operator
product expansion for the vacuum polarization, the first non trivial
operator to be considered is ${\rm F}_{\mu\nu}{\rm F}_{\mu\nu}$. It
has dimension four and thus gives a contribution proportional to
$q^{-4}$ that should compensate the imaginary part of the vacuum
polarization arising from the infrared renormalon at $b\beta_2=-2$
\cite{Muellernp}. We want to extract this contribution adding to the
action the multilocal operator dual to ${\rm F}^2$: 
\beq
\Delta S_2=i\frac{C}{12}\Lambda^4\int d^4x{\rm F}_{\mu\nu}(x)\int d^4y{\rm F}_{\mu\nu}(y)\label{deltas2}
\eeq 
The constant $C$ is fixed by the condition that the insertion of
$\Delta S_2$ on $\Pi_0$ cancel the imaginary part of $\Pi_1$.

In general the insertion of $\Delta S_2$ on some correlation function
can be computed with the Schwinger operator method in the so-called
fixed-point gauge; the details of the calculation will be given in
the appendix. The result for the insertion of $\Delta S_2$ on
$\Pi_{0\,\mu\nu}$ is:
\beq
-i\frac{C}{N_f}\frac{2}{3}\frac{a^2}{q^4}\Lambda^4(q_\mu q_\nu -q^2g_{\mu\nu})
\eeq
Comparing this expression with eq.(\ref{impolir}), we find for the
constant $C$ the value:
\beq
C=\frac{9}{8a}\label{const}
\eeq
As a fundamental check for the consistence of the theory we try to
extract the same constant $C$ from the imaginary part of the other
functions. In fact, while there is no contribution of $\Delta S_2$ on
the fermion propagator (this is consistent with the form of
Im$\,\Sigma$), one can compute the insertion of $\Delta S_2$ on
$\Gamma_\mu$ at $q=0$. Referring again to the appendix for the details
of the calculation we find the following contribution:
\beq
i\frac{eC}{N_f}\Lambda^4\frac{2}{3}\frac{\pi a}{p^4}\left (\gamma_\mu-4\frac{p_\mu\hat{p}}{p^2}\right )\label{inser}
\eeq
This must be confronted with the term proportional to $\Lambda^4$ at
$q=0$ in Im$\,\Gamma_\mu$:
\beq
-\frac{1}{N_f}\frac{3}{4}\frac{\pi e}{p^4}\left (\gamma_\mu -4\frac{p_\mu\hat{p}}{p^2}\right )
\eeq 
Hence the same coefficient $C$ found in (\ref{const}) eliminates this
imaginary term.

For the sake of completeness we have extracted the coefficient $C$
directly from the imaginary part of  $\langle {\rm F}_{\mu\nu}{\rm
F}_{\mu\nu} \rangle$; from the diagram of fig.6 we find for the Borel
transform: 
\beq
\langle \widetilde{a {\rm F}^2}\rangle \mathop\simeq_{b\sim -6\pi}-
\frac{9}{8\pi}\frac{1}{b+6\pi}{\rm e}^{-5b/9\pi}\left
(\frac{1}{\mu^2}\right )^{b/3\pi}
\eeq
which implies:
\beq
{\rm Im}\;\langle {\rm F}_{\mu\nu}{\rm F}_{\mu\nu}\rangle =-
\frac{9}{8a} \Lambda^4+O(\Lambda^6)\label{imvac}
\eeq
Moreover the insertion of $\Delta S_2$ on  $\langle {\rm F}_{\mu\nu}{\rm
F}_{\mu\nu} \rangle$ is simply given by $C$ so that even this
imaginary part is cancelled with the choice (\ref{const}).

\section{Conclusions}
We have shown in a non trivial context that renormalon-type
singularities can be absorbed in an effective lagrangian of the form:
\beq
{\cal L}_{eff}={\cal L}+i \Delta {\cal L}
\eeq
$\Delta {\cal L}$ is uniqely determined by the request of having real
correlation functions in the euclidean region. Local counterterms
correspond to ultraviolet renormalons while multilocal counterterms to
infrared ones.

The imaginary part in ${\cal L}_{eff}$   restores unitarity
which is violated in the original theory for the presence of
renormalons.

The description in terms of an effective lagrangian makes sense only
if the resulting theory doesn't present non local singularities. We
have checked that this is the case for QED in the framework of the
$1/N_f$ expansion.

From the theoretical point of view it would be interesting to study
the global structure of ${\cal L}_{eff}$ and to understand how the
introduction of imaginary counterterms may influence the
renormalization of the theory.

\newpage

\appendix
\section{Background field technique}

We want to shortly review the background field formalism which enables
to compute  the correlation functions in an external field
\cite{Schwinger,Novikov}. 

The Schwinger method enables to enforce gauge-invariance at all stages of the calculation; it is based on the introduction of a basis of eigenvectors of the coordinate operator:
\beqn
X_\mu |x\rangle &=&x_\mu |x\rangle\\
\langle x|P_\mu|y\rangle &=&-i\frac{\partial}{\partial x_\mu}\delta (x-y) + e A_\mu (x)\delta (x-y)\nonumber
\eeqn 
In addition we perform our calculation in the so-called fixed-point
gauge \cite{Cronstrom}:
\beq
x_\mu A_\mu (x) =0
\eeq
In this gauge the vector potential is written in terms of the field
strenght and of its covariant derivatives calculated in $x=0$; in
momentum space we have:
\beq
A_\mu (k)=\int d^4z\, {\rm e}^{ikz}A_\mu (z)\simeq -i\frac{(2\pi )^4}{2}{\rm F}_{\rho\mu}(0)\frac{\partial}{\partial k_\rho}\delta^{(4)}(k)+\ldots\label{amuk}
\eeq 
From this formulas it is easy to derive the expression for the fermion propagator in an external field:
\beq
S(q)=\int dx \langle x|\frac{1}{\hat{P}+\hat{q}}|0\rangle =\frac{1}{\hat{q}}-\frac{1}{q^4}e\tilde{{\rm F}}_{\rho\sigma}q_\rho \gamma_\sigma\gamma_5+\ldots\label{propfer}
\eeq
where $\tilde{{\rm F}}_{\mu\nu}$ is the dual of ${\rm
F}_{\mu\nu}$.

\noindent 
We want to find the contributions of $\langle {\rm F}^2\rangle$ to the proper photon propagator and to the vertex function; for this purpose we must use the randomness of the field i.e.:
\beq
\langle {\rm F}_{\mu\nu}(0){\rm F}_{\rho\sigma}(0)\rangle =\frac{1}{12}(g_{\mu\rho}g_{\nu\sigma}-g_{\mu\sigma}g_{\nu\rho})\langle {\rm F}_{\alpha\beta}{\rm F}_{\alpha\beta}\rangle\label{random}
\eeq
The insertion of $\langle {\rm F}^2\rangle$ on $\Pi (q^2)$ has been
computed   many times in the literature (see for example
\cite{Shifman});  the expression for the photon propagator in an external field is given by (using (\ref{propfer}) and (\ref{random})):
\beqn
\Pi_{\mu\nu}(q)&=&ie^2\int d^4x\,{\rm e}^{iqx}\,{\rm Tr}\{\gamma_\mu S(x,0)\gamma_\nu S(0,x)\}=\\
&=&e^2\,(q_\mu q_\nu-q^2g_{\mu\nu})\left \{-\frac{1}{12\pi^2}{\rm log}q^2-\frac{e^2}{24\pi^2q^4}\langle {\rm F}^2\rangle\right\}\nonumber
\eeqn
The contribution of $\langle {\rm F}^2\rangle $ to the vertex function
at  zero momentum transfer comes from the insertion of a single
$A_\mu$ on  every propagator (see fig.7):
\beq
V_\mu^{\langle {\rm F}^2\rangle}(p)=-2e^3\,\int\frac{d^4k_1}{(2\pi)^4}\,\frac{d^4k_2}{(2\pi)^4}\,\frac{1}{\hat{p}+\hat{k_1}+\hat{k_2}}\,\hat{A}(k_2)\,\frac{1}{\hat{p}+\hat{k_1}}\,\gamma_\mu\,\frac{1}{\hat{p}+\hat{k_1}}\,\hat{A}(k_1)\,\frac{1}{\hat{p}}
\eeq
From (\ref{amuk}) and (\ref{random}) we have, considering the one-particle irreducible function:
\beq
\Gamma_\mu^{\langle {\rm
F}^2\rangle}(p)=\frac{1}{6}\,e^3\,\frac{\langle {\rm
F}^2\rangle}{p^4}\left [\gamma_\mu -4\frac{p_\mu\hat{p}}{p^2}\right ]
\eeq
which corresponds to eq.(\ref{inser})

\newpage
\noindent
{\bf FIGURE CAPTIONS}\\

\noindent
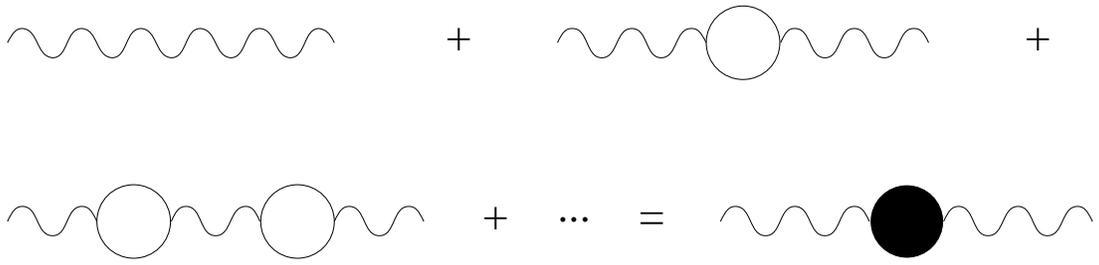
{\bf Fig 1.} The photon propagator at leading order.\\

\noindent
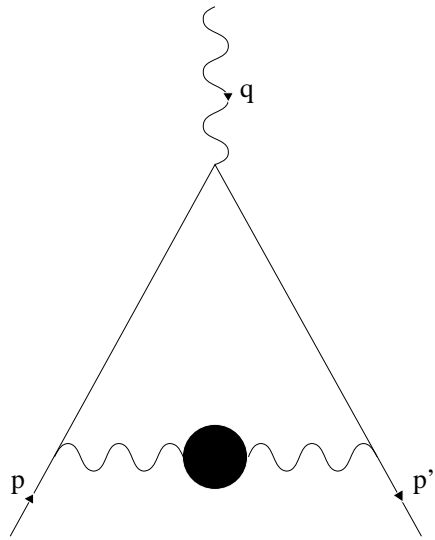
{\bf Fig 2.} The vertex function at order $1/N_f$.\\

\noindent
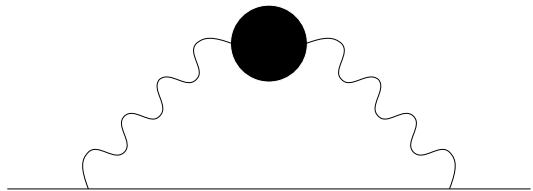
{\bf Fig 3.} The fermion propagator at order $1/N_f$.\\

\noindent
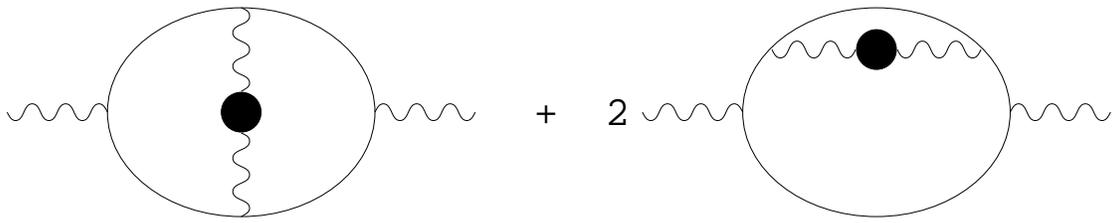
{\bf Fig 4.} The photon vacuum polarization at order $1/N_f$.\\

\noindent
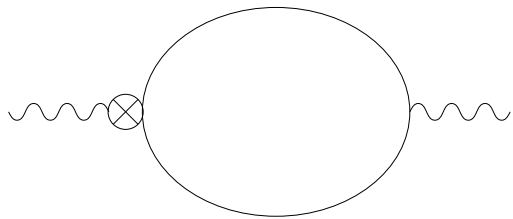
{\bf Fig 5.} Insertion of $\Delta {\cal L}_1$ (the crossed circle) on the
vacuum polarization.\\

\noindent
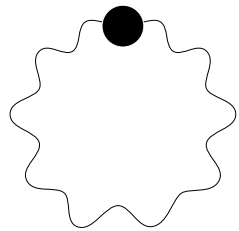
{\bf Fig 6.} Graph contributing to the vacuum expectation value of ${\rm
F}_{\mu\nu} {\rm F}_{\
mu\nu}$.\\

\noindent
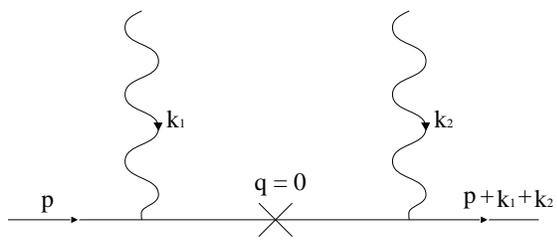
{\bf Fig 7.} Contribution of ${\rm F}^2$ to the vertex function at zero
momentum transfer.\\

\end{document}